\documentclass[12pt,preprint]{aastex}

\shorttitle{The softening phenomenon} \shortauthors{Qin}

\begin{document}

\title{The softening phenomenon due to the curvature effect: in the case of extremely short intrinsic
emission}
\author{Y.-P. Qin\altaffilmark{1,2}}

\altaffiltext{1}{Center for Astrophysics, Guangzhou University,
Guangzhou 510006, P. R. China; ypqin@gzhu.edu.cn}

\altaffiltext{2}{Physics Department, Guangxi University, Nanning
530004, P. R. China}

\begin{abstract}
Both the light curve and spectral evolution of the radiation from a
relativistic fireball with extremely short duration are studied, in
order to examine the curvature effect for different forms of the
radiation spectrum. Assuming a $\delta$ function emission we get
formulas that get rid of the impacts from the intrinsic emission
duration, applicable to any forms of spectrum. It shows that the
same form of spectrum could be observed at different times, with the
peak energy of the spectrum shifting from higher energy bands to
lower bands following $E_{peak}\propto t^{-1}$. When the emission is
early enough the $t^{2}f_{\nu }(t)$ form as a function of time will
possess exactly the same form that the intrinsic spectrum as a
function of frequency has. Assuming $f_\nu \propto \nu^{-\beta}
t^{-\alpha}$ one finds $\alpha=2+\beta$ which holds for any
intrinsic spectral forms. This relation will be broken down and
$\alpha > 2+\beta $ or $\alpha \gg 2+\beta $ will hold at much later
time when the angle between the moving direction of the emission
area and the line of sight is large. An intrinsic spectrum in the
form of the Band function is employed to display the light curve and
spectral evolution. Caused by the shifting of the Band function
spectrum, a temporal steep decay phase and a spectral softening
appear simultaneously. The softening phenomenon will appear at
different frequencies. It occurs earlier for higher frequencies and
later for lower frequencies. The terminating softening time
$t_{s,max}$ depends on the observation frequency, following
$t_{s,max}\propto \nu^{-1}$. This model predicts that the softening
duration would be linearly correlated with $t_{s,max}$; the observed
$\beta_{min}$ and $\beta_{max}$ are determined by the low and high
energy indexes of the Band function; both $\beta_{min}$ and
$\beta_{max}$ are independent of the observation frequency.
\end{abstract}


\section{Introduction}

The successful launch of the Swift satellite (Gehrels et al. 2004)
has made great advance for the observations of the X-ray afterglows
of gamma-ray bursts (GRBs) in the past few years (for a recent
review, see Zhang 2007). Among the many new findings obtained by the
Swift instruments, some emerge as a puzzling. One is the spectral
evolution which was detected in the tails of some bursts (Campana et
al. 2006; Ghisellini et al. 2006; Gehrels et al. 2006; Mangano et
al. 2007; Zhang et al. 2007a). The phenomenon has not been predicted
by the curvature effect nor by the standard external shock afterglow
model. Soon after that, Zhang et al. (2007b) performed a systematic
analysis on this issue and found that 33 of the 44 bursts with
bright steep decay tails show an obvious spectral evolution --- an
observed softening. This suggests that, the detected spectral
evolution is not a rare phenomenon, but instead, it is quite common
(at least detectable in the majority of GRBs that have bright steep
decay tails).

Several attempts have been made to interpret this softening
phenomenon. It was proposed that a central engine producing a soft
but decaying afterglow emission might be responsible for some of
these bursts (Campana et al. 2006; Fan et al. 2006; Zhang et al.
2007b). Some bursts with strong softening might be accounted for by
a cooling of the internal-shocked region (Zhang et al. 2007b). A
possible thermal component has also been tried. However, Yonetoku et
al. (2008) showed that introducing a thermal component is not
sufficient to explain all of the spectral softening, and thus
additional spectral evolution is required. In investigating the
origins for the spectral evolution of GRB 070616, Starling et al.
(2008) ruled out the possibility that a superposition of two
power-laws causes the evolution. In stead, they considered a
possibility of an additional component dominant during the late
prompt emission. They proposed that a combination of the spectral
evolution and the curvature effect may cause the observed steep
decay phase of the light curve. Another scenario is based on the
cannonball model of GRBs, which was suggested to be responsible for
both the temporal behavior and the spectral softening of the bursts
observed by Swift (Dado et al. 2008).

As shown in Zhang et al. (2007b), the spectral softening is
accompanied by a very steep decay phase. This phase is seen directly
following the prompt emission and is naturally (and generally)
regarded as the tail of the prompt emission (Tagliaferri et al.
2005; Barthelmy et al. 2005; Liang et al. 2006). The tail was
suspected to arise from the emission of the high latitude of the
fireball surface, which is often referred to as the curvature effect
(Kumar \& Panaitescu 2000; Dermer 2004; Dyks et al. 2005; Liang et
al. 2006; Panaitescu et al. 2006; Zhang et al. 2006, 2007b; Butler
\& Kocevski 2007; Qin 2008; Starling et al. 2008). A full
consideration of the curvature effect includes the delay of time and
the shifting of the intrinsic spectrum as well as other relevant
factors of an expanding fireball (for detailed explanation and
analysis, see Qin 2002; Qin et al. 2004, 2006; Qin 2008). The effect
has been intensively studied in the prompt gamma-ray phase, such as
the profile of the light curve of pulses, the spectral lags, the
power-law relation between the pulse width and energy, the evolution
of the hardness ratio and the evolution of the peak energy (Fenimore
et al. 1996; Sari \& Piran 1997; Qin 2002; Ryde \& Petrosian 2002;
Kocevski et al. 2003; Qin \& Lu 2005; Shen et al. 2005; Lu et al.
2006, 2007; Peng et al. 2006; Qin et al. 2004, 2005, 2006; Jia
2008).

In a recent investigation, Butler \& Kocevski (2007) concluded that the
early emission in $>90\%$ of early afterglows has a characteristic $\nu
f_{\nu}$ spectral energy $E_{peak}$, which likely evolves from the $\gamma$%
-rays through the soft X-ray band on timescales of $10^2 - 10^4$ s
after the GRB. Many careful analyses revealed that there do exist
some bursts with their peak energy $E_{peak}$ decreasing from a
higher band to a much lower band. These bursts include: GRB 060124,
from 108 keV to 1.3 keV (Butler \& Kocevski 2007); GRB 060614, from
8.6 keV to 1.1 keV (Butler \& Kocevski 2007; Mangano et al. 2007);
GRB 060904A, from 163 keV to 2.28 keV (Yonetoku et al. 2008); GRB
061121, from 270 keV to 0.95 keV (Butler \& Kocevski 2007); GRB
070616, from 135 keV to $\sim$4 keV (Starling et al. 2008). For GRB
070616, the spectral softening evolution was observed even in the
prompt emission phase: its duration is $T_{90}=402.4s$, while the
softening starts from 285s and extends to 1200s (Starling et al.
2008). Among them, GRB 060614, GRB 060904A and GRB 061121 are
members of the Zhang et al. (2007b)'s sample. Although the curvature
effect was rejected to interpret the softening by some authors,
Starling et al. (2008) insisted that the observed shifting of the
peak energy is in agreement with what expected by the curvature
effect: the peak energy of the Band function spectrum passes through
the $\gamma$-ray band at a relatively early time, while it passes
through the X-ray band at a later time due to the high latitude
emission. They proposed that both the curvature effect and a strong
spectral evolution cause the steep decline in flux. Based on the
explicit illustration of the evolution of the whole spectral form in
Butler \& Kocevski (2007), we suspect that the curvature effect
alone might be responsible for both the spectral softening and the
accompanied steep decay light curve.

To reveal the pure curvature effect and get rid of the possible
impacts from the emission duration, we concern in this paper only
extremely short intrinsic emission. Focusing on this emission has
two advantages. The first is that the formulas become very simple,
and the second is that many key characteristics of the effect can be
plainly illustrated. The paper is organized as follows. In Section
2, we present basic formulas of the full curvature effect,
applicable to any temporal and spectral forms of emission. We
discuss light curves and spectral evolution arising from an
extremely short emission in Section 3. In Section 4, we assume an
intrinsic spectrum in the form of the Band function and illustrate
the corresponding light curve and spectral evolution in detail.
Parameters that affect the results are discussed in Section 5. In
Section 6, we apply the model to the XRT band and also to a Swift
burst. Discussion and conclusions are presented in Section 7.

\section{Equations of flux densities influenced by the curvature effect}

In the following, we study the emission from an expanding fireball
shell. Suppose the emission occurs over the fireball area confined
by $\theta _{\min }\leq \theta \leq \theta _{\max }$, where $\theta
$ is the angle between the normal of the area concerned with respect
to the line of sight (which is also the angle of the moving
direction of the emitting region with respect to the direction to
the observer), and within proper time interval $t_{0,\min }\leq
t_{0}\leq t_{0,\max }$. Let us consider the following situation: the
Lorentz factor of the emitting shell is constant, the energy range
of the emission is unlimited, and the intrinsic radiation intensity
is independent of direction. Basic formulas of the flux density
which is expected by a distant observer measured at laboratory time
$t_{ob}$ and other relevant quantities for this simple situation are
presented in Qin (2008; see equations 1-5 there).

To meet and/or approximate the conventional definition of
observation time, we assign (see also Qin 2008)
\begin{equation}
t\equiv t_{ob}-t_{c}+R_{c}/v-D/c,
\end{equation}%
where $t_{c}$ as a time constant is defined in the observer frame,
$D$ is the distance of the fireball to the observer, $v$ is the
speed of the shell, $R_{c}$ is the radius of the shell measured at
$t_{c}$ by local observers who are stationary in the explosion area.
Equation (1) is a definition of observation time. With this
definition of time, the basic formulas can be written as
\begin{equation}
f_{\nu }(t)=\frac{2\pi c^{2}}{D^{2}(\Gamma v/c)^{2}t^{2}}\int_{\widetilde{t%
}_{0,\min }}^{\widetilde{t}_{0,\max }}I_{0,\nu }(t_{0},\nu
_{0})[R_{c}/c+(t_{0}-t_{0,c})\Gamma v/c]^{2}[(t_{0}-t_{0,c})\Gamma
+R_{c}/v-t]dt_{0},
\end{equation}%
with%
\begin{equation}
\widetilde{t}_{0,\min }=\max \{t_{0,\min
},\frac{t-R_{c}/v+(R_{c}/c)\cos \theta _{\max }}{[1-(v/c)\cos \theta
_{\max }]\Gamma }+t_{0,c}\},
\end{equation}%
\begin{equation}
\widetilde{t}_{0,\max }=\min \{t_{0,\max
},\frac{t-R_{c}/v+(R_{c}/c)\cos \theta _{\min }}{[1-(v/c)\cos \theta
_{\min }]\Gamma }+t_{0,c}\},
\end{equation}%
\begin{equation}
\nu _{0}=\frac{t}{R_{c}/v+(t_{0}-t_{0,c})\Gamma }\Gamma \nu ,
\end{equation}%
and%
\begin{equation}
\begin{array}{l}
\lbrack 1-(v/c)\cos \theta _{\min }][(t_{0,\min }-t_{0,c})\Gamma
+R_{c}/v]\leq t \\
\leq \lbrack 1-(v/c)\cos \theta _{\max }][(t_{0,\max }-t_{0,c})\Gamma
+R_{c}/v]%
\end{array}%
,
\end{equation}%
where $t_{0,c}$ denotes the moment of $t_{c}$, measured by a
co-moving observer; $I_{0,\nu }(t_{0},\nu _{0})$ is the intrinsic
radiation intensity; $\Gamma =1/\sqrt{1-v^{2}/c^{2}}$. Equations
(2)-(6) are more general than equations (8)-(12) in Qin (2008). The
former can be applied to any forms of the intrinsic spectrum, while
the later are applicable only in the case of a single power-law
spectrum. According to equation (1), referring also to equation (8)
in Qin et al. (2004), $t=0$ corresponds to the moment of the
emission that occurs at the spot of the explosion (say, $ R_{c}=0$).
Or precisely, $t=0$ is the moment when photons emitted from $
R_{c}=0$ reach the observer. As explained in Qin (2008), observation
time $t$ approximates the time defined by the trigger time (e.g.,
$t=t_{ob}-t_{ob,trig}$). As long as the Lorentz factor is large
enough and the trigger event is early enough, the offset between the
two definitions of observation time would be very small (Qin 2008).

\section{Light curves and the spectral evolution of the fireball arising
from an extremely short intrinsic emission}

The simplest emission is a $\delta$ function emission which can
always simplify the equations concerned. Perhaps the most important
reason for considering a $\delta$ function emission is that effects
arising from the duration of real intrinsic emission will be omitted
and therefore those merely coming from the expanding motion of the
fireball surface will be clearly seen (Qin 2008). In practical
situation, when an emission is extremely short, one could regard it
as a $\delta$ function emission. In order to reveal the main
properties of the curvature effect in the case of
X-ray afterglow, we consider only this kind of radiation and hence assume a $%
\delta$ function emission through out this paper.

Let the concerned intrinsic emission be
\begin{equation}
I_{0,\nu }(t_{0},\nu _{0})=I_{0,0}\delta
(\frac{t_{0}}{t_{0,0}}-\frac{t_{0,c}}{t_{0,0}})g_{\nu }(\nu _{0}),
\end{equation}%
where $t_{0,0}>0$ is any assigned time constant (e.g.,
$t_{0,0}=1s$), $I_{0,0}$ is a constant in units of $erg\cdot cm^{-2}
s^{-1} Hz^{-1}$, and $g_{\nu }(\nu _{0})$ is the intrinsic spectrum
of the emission in a dimensionless form. We consider the emission
from the whole fireball surface and take $\theta _{\min }=0$ and
$\theta _{\max }=\pi /2$. In this situation, one gets from (6) that
(see also Qin 2008)
\begin{equation}
(1-v/c)R_{c}/v\leq t\leq R_{c}/v.
\end{equation}%
Within this observation time interval, equation (2) becomes
\begin{equation}
f_{\nu }(t)=\frac{2\pi I_{0,0}t_{0,0}R_{c}^2}{D^{2}(\Gamma
v/c)^{2}}\frac{g_{\nu }[\nu _{0}(t_0 = t_{0,c})](R_{c}/v-t)}{t^{2}}.
\end{equation}%
According to equation (5), $\nu _{0}$ is related to $t$ and $\nu$ by
\begin{equation}
\nu _{0}=\frac{\Gamma v}{R_{c}}t\nu .
\end{equation}
Inserting equation (10) into equation (9) comes to a plain result:
\begin{equation}
f_{\nu }(t)=\frac{2\pi I_{0,0}t_{0,0}R_{c}^2}{D^{2}(\Gamma v/c)^{2}}%
(R_{c}/v-t)g_{\nu }(\frac{\Gamma v}{R_{c}}t\nu )t^{-2}.
\end{equation}

A straightforward consequence of equation (11) comes from the
situation when $g_{\nu }(\nu_0)\propto \nu_0^{-\beta}$, that gives
rise to $f_{\nu }(t) \propto (R_{c}/v-t)t^{-2-\beta}\nu^{-\beta}$
(see Qin 2008). When the event occurs early enough, the $R_{c}/v-t$
term will become constant and then we come to the well-known form of
flux density $f_{\nu }(t) \propto t^{-2-\beta}\nu^{-\beta}$
(Fenimore et al. 1996; Kumar \& Panaitescu 2000; Qin 2008).

Several conclusions are reached from equation (11). a) For a certain
observation time satisfying equation (8), the spectrum observed is
merely a shifted intrinsic one. The shifting factor is $\Gamma
vt/R_{c}$. The peak energy will decline following
$E_{peak}=E_{peak,0}R_{c}/ \Gamma vt$ (namely, $E_{peak}\propto
t^{-1}$). b) For a certain observation frequency, the light curve
observed depends entirely on the emission spectrum. In the case when
$R_{c}/v\gg t$, the form $t^{2}f_{\nu }(t)$, $\propto g_{\nu
}(\frac{\Gamma v\nu }{R_{c}}t)$, as a function of time takes the
same form that the intrinsic spectrum as a function of frequency
has. Or, from $t^{2}f_{\nu }(t)|_{\nu=const}$, when replacing $t$
with variable $\nu$ and multiplying it with a constant to alter its
dimension one will get exactly the intrinsic spectral form (this
might be useful in checking the curvature effect in further
investigations). This is plain in the pure power-law emission where
$f_{\nu }(t) \propto t^{-2-\beta}\nu^{-\beta}$ in the case
$R_{c}/v\gg t$. But the conclusion holds for any intrinsic spectral
forms, which is unaware previously. As explained in Qin (2008), the
term $R_{c}/v-t$ reflects the projected factor of the infinitesimal
fireball surface area in the angle concerned to the distant
observer, known as $\cos \theta $ (in fact, $R_{c}/v-t$ is $\cos
\theta $ multiplying a constant; see Appendix A). When the emission
area is close to the line of sight region, condition $R_{c}/v\gg t$
(i.e., $\cos \theta \sim 1$) can easily be satisfied, while when the
angle between the moving direction of the emission region and the
line of sight is large enough the term $R_{c}/v-t$ will be
important. At a much later time, the emission area would be close to
that of $\theta =\pi /2$, and the term $R_{c}/v-t$ and then the flux
will approach to zero (see Appendix A and Appendix B). c) Under the
condition of $R_{c}/v\gg t$, which will hold at an earlier time, the
temporal power law index and the spectral power law index will be
well related. For a given time $t$ and a given frequency $\nu $, we
assign the flux as $f_{\nu }(t)=f_{0}t^{-\alpha }$ and assign the
intrinsic spectrum as $g_{\nu }(\frac{\Gamma v}{R_{c}}t\nu
)=g_{0}\cdot (\frac{\Gamma v }{R_{c}}t\nu )^{-\beta }$. These can
easily be satisfied when one carefully chooses $f_{0}$, $\alpha $,
$g_{0}$ and $\beta $ in the vicinity of $t$ (e.g., by performing a
fit). Inserting $g_{\nu }(\frac{\Gamma v}{R_{c}}t\nu )=g_{0}\cdot
(\frac{\Gamma v }{R_{c}}t\nu )^{-\beta }$ into $f_{\nu
}(t)=f_{0}t^{-\alpha }$ under the $R_{c}/v\gg t$ condition one will
come to the well-known $\alpha =2+\beta $ relation from equation
(11). Note that this relation will hold for any intrinsic spectral
forms as long as the condition $R_{c}/v\gg t$ is satisfied. When the
angle $\theta$ is large enough (say, when $R_c/v-t \ll R_{c}/c$; see
Appendix A) the $\alpha =2+\beta $ relation will be broken down
since the influence of the $R_c/v-t$ term on the differential of the
light curve (which is associated with $\alpha$) will no more be
ignored.

\section{In the case of the Band function spectrum}

According to the above analysis, for an extremely short emission
burst, its observed spectrum and its intrinsic spectrum take the
same form. It is known that the observed spectra of most GRBs
possess the Band function form (Band et al. 1993). Since emission
from some of such bursts might be extremely short, it is likely that
the intrinsic emission of some bursts takes the Band function form.
In this section, we consider an intrinsic Band function emission and
assume (Band et al. 1993)
\begin{equation}
g_{\nu }(\nu _{0})=\{%
\begin{array}{c}
(\frac{\nu _{0}}{\nu _{0,p}})^{1+\alpha _{B,0}}\exp [-(2+\alpha _{B,0})\frac{%
\nu _{0}}{\nu _{0,p}}]\qquad \qquad \qquad \qquad (\frac{\nu _{0}}{\nu _{0,p}%
}<\frac{\alpha _{B,0}-\beta _{B,0}}{2+\alpha _{B,0}}) \\
(\frac{\alpha _{B,0}-\beta _{B,0}}{2+\alpha _{B,0}})^{\alpha
_{B,0}-\beta _{B,0}}\exp (\beta _{B,0}-\alpha _{B,0})(\frac{\nu
_{0}}{\nu _{0,p}})^{1+\beta
_{B,0}}\qquad (\frac{\alpha _{B,0}-\beta _{B,0}}{2+\alpha _{B,0}}\leq \frac{%
\nu _{0}}{\nu _{0,p}})%
\end{array}%
,
\end{equation}%
where $\nu _{0,p}$, $\alpha _{B,0}$, and $\beta _{B,0}$ are
constants.

With this spectral form, we can produce the light curve and the
spectral evolution using equation (11). In the following analysis we
take $\Gamma =100 $, $R_{c}=10^{15}cm$, $1+\alpha _{B,0}=-0.5$,
$1+\beta _{B,0}=-2.5$, and $h\nu _{0,p}=E_{0,p}=1keV$ as a primary
set of parameters. In the consequent analysis below they will be
replaced one by one to reveal their influences on the light curve
and the spectral evolution. We assign $F_0 \equiv 2\pi
I_{0,0}R_{c}^2/D^{2}(\Gamma v/c)^{2}$ and $t_{0,0}=1s$. The flux
density will be calculated in units of $F_0$ through out this paper.

\subsection{Expected at the 1 keV observation frequency}

First, let us explore the spectral evolution at $E=1keV$ by assuming
a power
law of flux within a limited (or narrow) band including $E=1keV$:%
\begin{equation}
f_{\nu }(t)=I(t)\nu ^{-\beta }.
\end{equation}

\begin{figure}[tbp]
\begin{center}
\includegraphics[width=5in,angle=0]{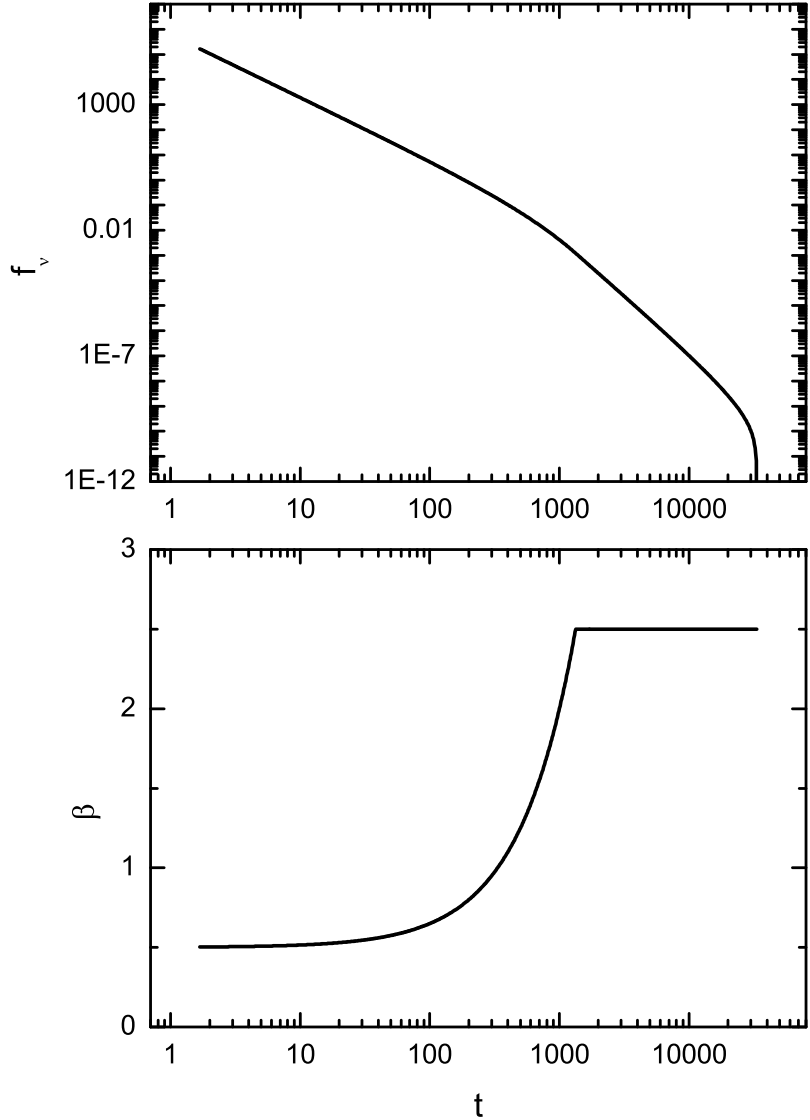}
\end{center}
\caption{The light curve (the upper panel) and spectral evolution
(the lower panel) of a $\delta$ function emission from an expanding
fireball, expected at 1 keV observation frequency. Observation time
$t$ is defined by equation (1) and in units of $s$. The flux density
is calculated with equation (11) and in units of $F_0$ [$F_0 \equiv
2\pi I_{0,0}R_{c}^2/D^{2}(\Gamma v/c)^{2}$], while the intrinsic
spectrum is determined by equation (12). The adopted parameters
include: $\Gamma =100 $, $R_{c}=10^{15}cm$, $1+\alpha _{B,0}=-0.5$,
$1+\beta _{B,0}=-2.5$, and $h\nu _{0,p}=1keV$. Equation (13) is used
to defined and evaluate the spectral index $\beta$.} \label{Fig.
1_01}
\end{figure}

The resulting light curve and spectral evolution are displayed in
Fig. 1. A temporal steep decay phase and a spectral softening are
observed within the range of $1-10^{3}$ s, which occur
simultaneously. Determined or influenced by the intrinsic spectral
form, the light curve decays in a milder  manner at an earlier time
(influenced by the lower energy index of the Band function) and then
turns to be steeper at a much later time (influenced by the higher
energy index of the Band function). Connecting these two segments is
a breaking feature which appears when the peak energy of the
spectrum passes through the observation band (here, the 1 keV
observation frequency). This feature is also viewable in Fig. 1 of
Kumar \& Panaitescu (2000) (at $\sim$ 400 s), and it is interpreted
as due to the passing through the observation band as well. We
observe that, at about 1400 s, the softening stops and the spectral
index becomes constant. This occurs after the peak energy has
sufficiently passing through the observation band. At about 30000 s,
the light curve ends with a cutoff tail which is determined by the
term $R_{c}/v-t$. As mentioned above, the term $R_{c}/v-t$ comes
from the projected factor of the infinitesimal fireball surface area
in the angle concerned (say, $\theta $) to the distant observer,
known as $\cos \theta $. As the angle between the moving direction
of the dominant emission area and the line of sight becomes larger,
the $\cos \theta $ term becomes smaller. When $\theta $ approaches
to $\pi /2$ (the edge of the half fireball surface that faces the
observer), $\cos \theta $ approaches to zero and then the tail comes
into being (it is expectable that any emission from a fireball must
be limited due to its limited size).

The observed XRT light curves are ranging from $60 s$ to $1\times
10^7 s$ and from $5\times 10^{-15} erg\cdot cm^{-2} s^{-1}$ to
$8\times 10^{-8} erg\cdot cm^{-2} s^{-1}$. From the data of Fig. 1
we find that, at $60s$ $f_{\nu}\sim 20 F_0$, at $1000s$ $f_{\nu}\sim
4\times 10^{-3} F_0$ (where the breaking feature appears), and at
$30000s$ $f_{\nu}\sim 1\times 10^{-10} F_0$ (where the light curve
cutoff tail, or the broken down feature, emerges). For the adopted
parameter set, the flux at $1000s$ is smaller than that at $60s$
about 4 orders of magnitude, and then the breaking feature is
reasonably expectable. The broken down feature cannot be observable
since the flux associated with it is smaller than that at $60s$
about 11 orders of magnitude. However, this does not mean that this
feature will never be observable since the magnitude of the flux
associated with it depends strongly on the fireball radius, the
intrinsic peak energy, and the energy indexes of the Band function
spectrum. For example, for a much smaller high energy index [say,
when $-(1+\beta _{B,0})$ is much smaller] the problem will be
significantly eased, and for a smaller fireball radius the interval
between the observable start time of the XRT light curve and the
broken down feature will be much shorter and the difference between
their flux magnitudes will be much smaller (see Fig. 6 and also the
discussion below).

\begin{figure}[tbp]
\begin{center}
\includegraphics[width=5in,angle=0]{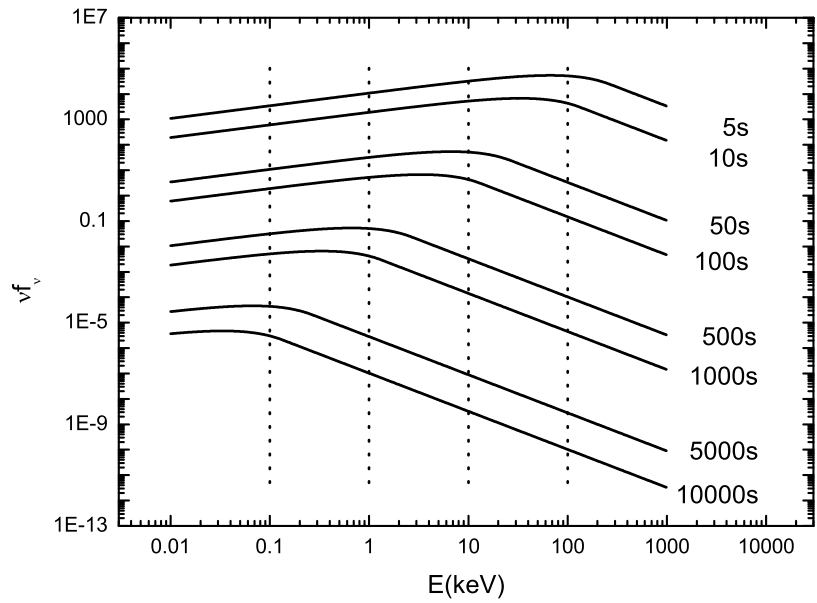}
\end{center}
\caption{The $\nu f_{\nu}$ (in units of $F_0 keV$) spectra of the
emission considered in Fig. 1, at different observation times. The
four dotted lines represent the 0.1, 1, 10, and 100 keV frequencies
respectively.} \label{Fig. 1_02}
\end{figure}

Development of the whole spectrum over the same period concerned is
displayed in Fig. 2, where the $\nu f_{\nu }$ curves at 5, 10, 50,
100, 500, 1000, 5000, and 10000 s, spanning from 0.01 to 1000 keV,
are presented. This figure plainly illustrates that, due to the
contribution of the high latitude emission (where angle $\theta$
becomes larger and larger), it is indeed that the shifting of the
Band function spectrum causes the softening observed in Fig. 1. When
the peak energy has passed through the adopted bandpass (say, the 1
keV observation frequency), the higher energy power law portion in
the intrinsic Band function spectrum gradually dominates the
emission. After $\sim$ 1000 s, the expected flux density (at
$h\nu=1keV$) is mainly contributed by this emission (i.e., the $\nu
_{0}^{1+\beta _{B,0}}$ portion emission), and then the
$t^{-2-\beta}\nu^{-\beta}$ curve comes into being (here
$-\beta=1+\beta _{B,0}$). In this period, the spectral index $\beta$
is of course constant (see Fig. 1). As mentioned above, the
$t^{-2-\beta}\nu^{-\beta}$ curve is deduced by assuming an intrinsic
emission with a power law spectrum, emitted from an expanding
fireball (Fenimore et al. 1996; Kumar \& Panaitescu 2000; Qin 2008),
which is a consequence of equation (11).

\begin{figure}[tbp]
\begin{center}
\includegraphics[width=5in,angle=0]{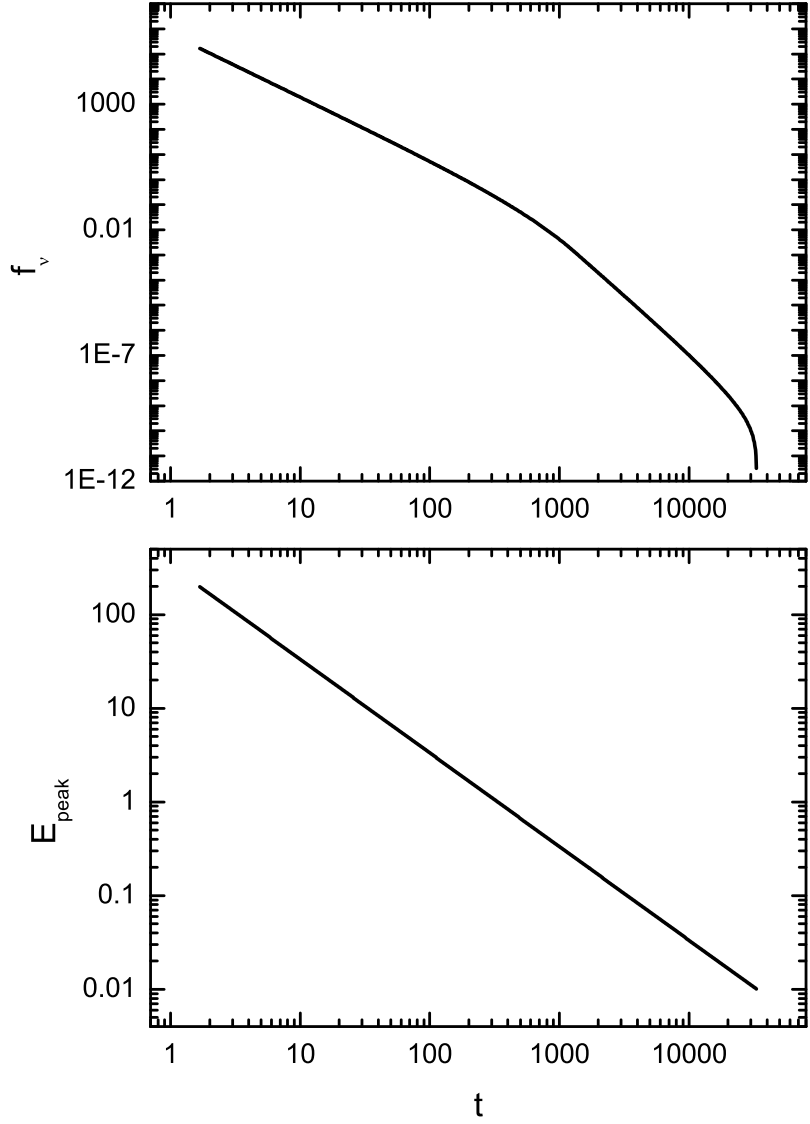}
\end{center}
\caption{Evolution of the peak energy $E_{peak}$ (the lower panel)
of the observed spectrum of the emission considered in Fig. 1, where
the light curve (the upper panel) is the same in Fig. 1.}
\label{Fig. 1_03}
\end{figure}

The evolution of the peak energy is shown in Fig. 3, where the
$E_{peak}\propto t^{-1}$ law is visible.

\begin{figure}[tbp]
\begin{center}
\includegraphics[width=5in,angle=0]{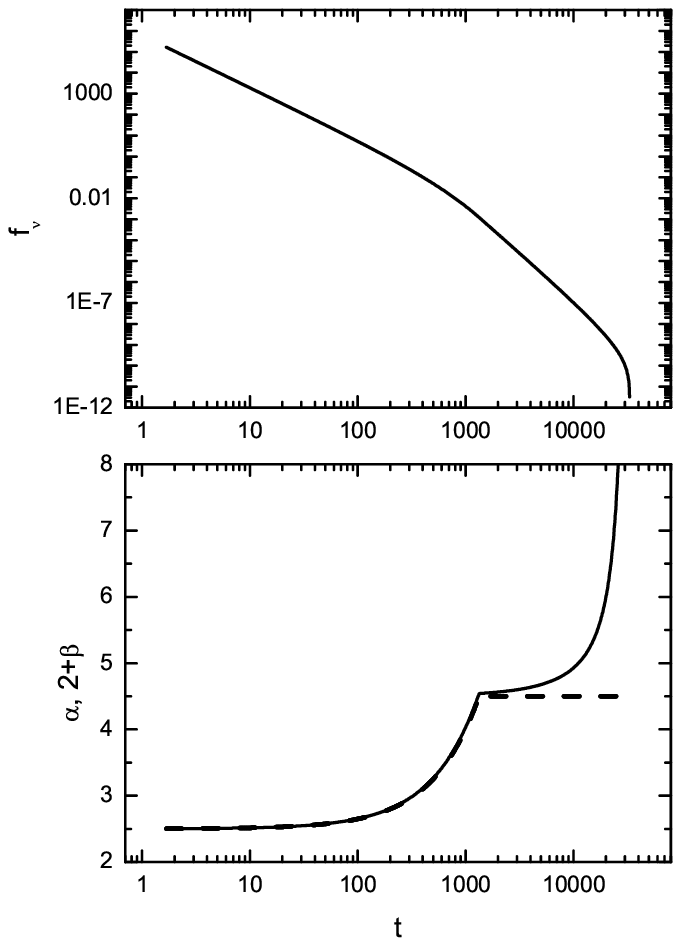}
\end{center}
\caption{Comparison of the temporal index $\alpha$ (the solid line
in the lower panel) and spectral index $\beta$ (presented in
$2+\beta$) (the dash line in the lower panel) of the emission
considered in Fig. 1, where the light curve (the upper panel) is the
same in Fig. 1.} \label{Fig. 1_04}
\end{figure}

The $\alpha =2+\beta $ relation discussed in last section can be
directly checked by plotting and comparing the $\alpha$ vs. $t$
curve and the $2+\beta$ vs. $t$ curve. This is shown in Fig. 4. When
the observation time is not too late (say, $< 2000$ s), the $\alpha
=2+\beta $ relation firmly stands. As expected, at later times the
$\alpha =2+\beta $ relation is broken down as the $\cos \theta $
term (the $R_{c}/v-t$ term) becomes important. This happens at
$\sim$ 2000 s. After that, the $\cos \theta $ term dominates the
flux, and then the temporal index $\alpha$ rapidly increases.

\subsection{Expected at other observation frequencies}

Next, let us consider the light curve and spectral evolution
expected at other observation frequencies. One finds from Fig. 2
that the peak energy passes through 0.1 keV frequency at a much
later time while it passes through 100 keV at a very early time.
This suggests that for the adopted set of parameters, the softening
would appear at lower and very higher energy bands as well, but the
corresponding times would be very different. Displayed in Fig. 5 are
the light curves and the spectral evolution expected at frequencies
0.1, 1, 10 and 100 keV respectively. The spectral index $\beta$
reaches its maximum at $\sim$ $1.3\times 10^4$, $1.3\times 10^3$,
$1.3\times 10^2$ and $13$ s, for frequencies 0.1, 1, 10 and 100 keV,
respectively.

\begin{figure}[tbp]
\begin{center}
\includegraphics[width=5in,angle=0]{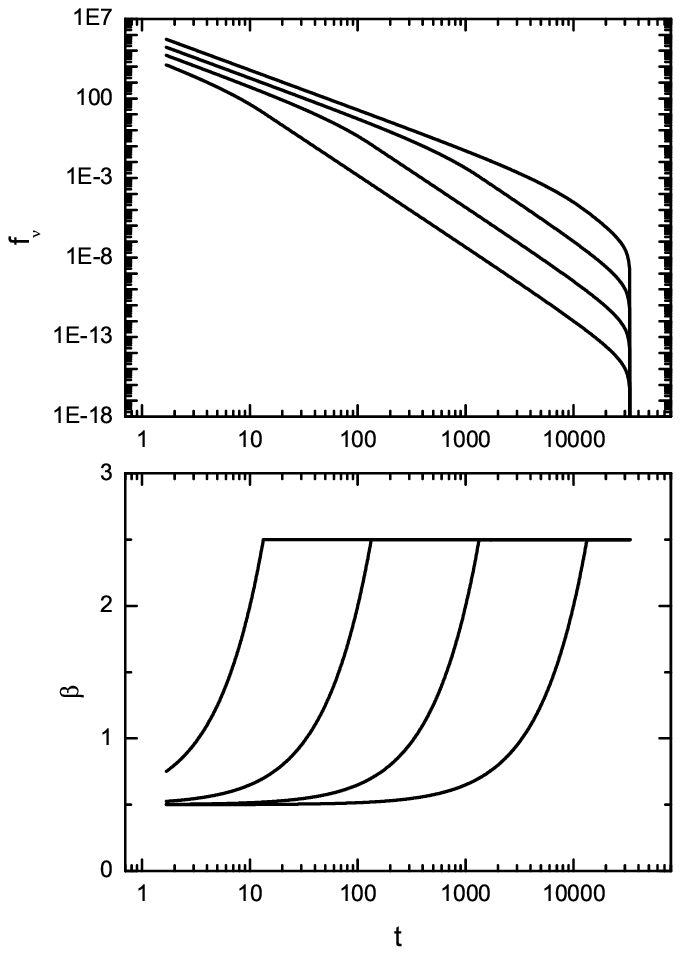}
\end{center}
\caption{Comparison of the light curve (the upper panel; the units
are the same as in Fig. 1) and the spectral evolution (the lower
panel) of the same emission considered in Fig. 1, expected at
different observation frequencies. In the lower panel, solid lines
from the left to the right correspond to observation frequencies
100, 10, 1, and 0.1 keV respectively, whilst in the upper panel,
solid lines from the bottom to the top are associated with 100, 10,
1, and 0.1 keV respectively.} \label{Fig. 1}
\end{figure}

This analysis leads to the following conclusions: the softening due
to the curvature effect will appear at different frequencies; it
occurs earlier for higher frequencies and later for lower
frequencies; the terminating softening time (defined as the time
when the spectral index $\beta$ reaches its maximum), $t_{s,max}$,
depends strictly on the observation frequency, and it follows the
$t_{s,max}\propto \nu^{-1}$ law, where $\nu$ is the observation
frequency. Note that the $t_{s,max}\propto \nu^{-1}$ law can also be
deduced from equation (10) and/or equation (11). This prediction
holds as long as the intrinsic emission is short enough.

Indeed, the spectral softening evolution was observed in the prompt
emission of GRB 070616, starting from 285s and extending to 1200s
(Starling et al. 2008). This favors our new finding that the
softening is also expectable in higher energy bands.

\section{Parameters that affect the results}

The light curve and spectral evolution discussed in last section are
produced by adopting a certain set of parameters. Here we
investigate how these parameters affect the results.

\subsection{Lorentz factor, fireball radius and peak energy of the intrinsic
spectrum}

The softening time scale exhibited in Fig. 1 must be affected by the
Lorentz factor, fireball radius and peak energy of the intrinsic
spectrum. We repeat the above analysis by replacing $\Gamma =100$,
$R_{c}=10^{15}cm$ and $h\nu _{0,p}=E_{0,p}=1keV$ with $\Gamma =10$,
$R_{c}=10^{14}cm$ and $h\nu _{0,p}=E_{0,p}=0.1keV$ one by one. The
results are displayed in Fig. 6.

\begin{figure}[tbp]
\begin{center}
\includegraphics[width=5in,angle=0]{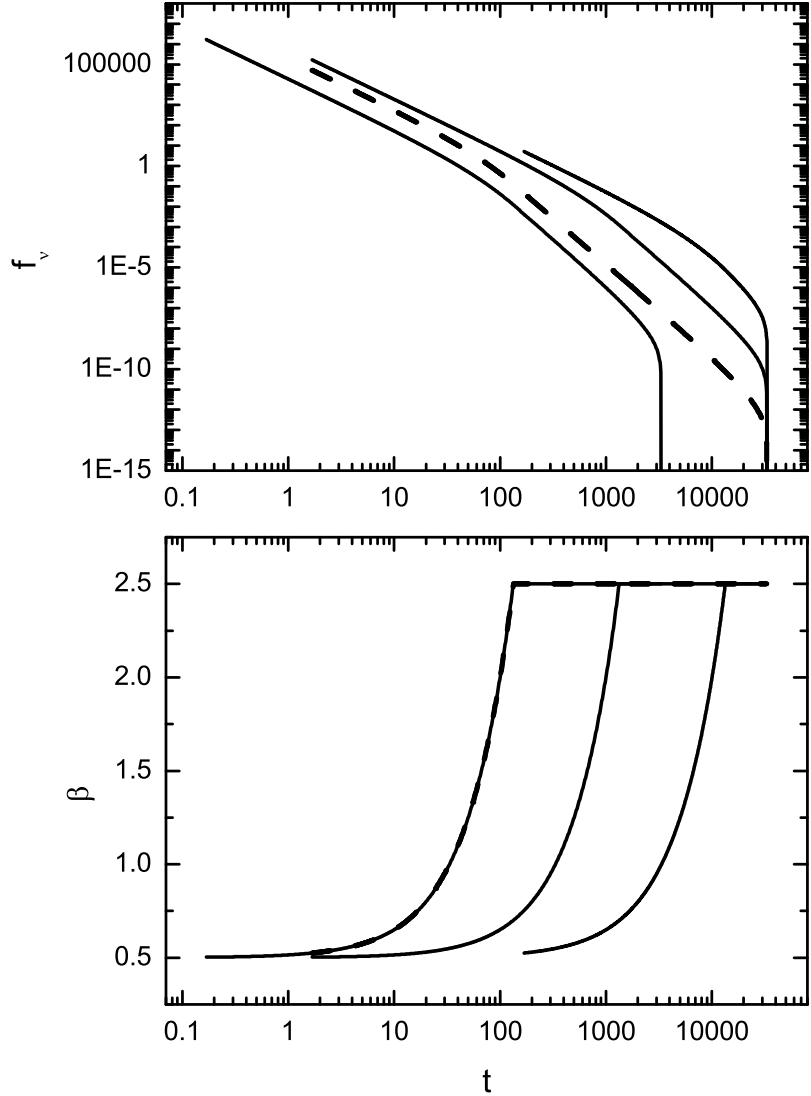}
\end{center}
\caption{The light curve (the upper panel) and spectral evolution
(the lower panel) of the emission considered in Fig. 1 (in the same
units), produced by replacing the parameters adopted there. In the
lower panel, solid lines from the left to the right correspond to
the cases of replacing $R_{c}=10^{15}cm$ with $R_{c}=10^{14}cm$,
without replacement, and replacing $\Gamma =100 $ with $\Gamma =10
$, respectively, and the dash line corresponds to the case of
replacing $h\nu _{0,p}=1keV$ with $h\nu _{0,p}=0.1keV$. In the upper
panel, solid lines from the bottom to the top are associated with
the cases of replacing $R_{c}=10^{15}cm$ with $R_{c}=10^{14}cm$,
without replacement, and replacing $\Gamma =100 $ with $\Gamma =10
$, respectively, and the dash line corresponds to the case of
replacing $h\nu _{0,p}=1keV$ with $h\nu _{0,p}=0.1keV$.} \label{Fig.
1_05}
\end{figure}

We find that a smaller Lorentz factor extend both the light curve
(marked by its breaking feature which corresponds to the moment when
the peak energy passes though the observation frequency) and the
spectral softening to a larger time scale. This must be resulted
from the less contraction of time. The cutoff tail of the steep
decay phase remains in the same time position. This is not surprise
since the tail is associated with the fireball radius, entirely
independent of the Lorentz factor (see Qin 2008).

As expected, in the case of $R_{c}=10^{14}cm$, the softening and the
cutoff tail (as well as the breaking feature) shift to earlier time.
The softening appears as early as $<3$ s and ends as early as
$\sim100$ s, and the cutoff tail appears at $\sim 3000$ s. At $60s$
we find $f_{\nu}\sim 0.3 F_0$, and at $3000s$ (where the light curve
cutoff tail appears) $f_{\nu}\sim 1\times 10^{-9} F_0$. For this
adopted parameter set, the flux at $3000s$ is smaller than that at
$60s$ about 8 orders of magnitude, which is only 1 order of
magnitude smaller than the usual observation magnitude range (7
orders). The undetectable problem raised above is now largely eased.

In the case of $h\nu _{0,p}=E_{0,p}=0.1keV$, the spectral softening
curve is almost overlapped with that of $ R_{c}=10^{14}cm$. Compared
with the case of $h\nu _{0,p}=E_{0,p}=1keV$, the breaking feature
shifts to earlier time, whilst the cutoff tail remains in the same
time position.

We notice from Fig. 6 together with Fig. 5 that the softening
process, associated with various parameters, spans a time scale
comparable to the terminating softening time $t_{s,max}$. In terms
of mathematics, this is due to the fact that, relative to the moment
$t=0$, the start time of the softening is much smaller (more than
one order of magnitude smaller) than $t_{s,max}$. We come to this
conclusion: the time interval of the softening is in the same order
of magnitude of the terminating softening time, and the two
quantities much be linearly correlated. We suspect that it is the
geometric property of the fireball surface that gives rise to this
relation.

\subsection{High and low energy spectral indexes}

High and low energy spectral indexes of the Band function must have
influences on the spectral softening. We replace $1+\alpha
_{B,0}=-0.5$ with $1+\alpha _{B,0}=0$ and $1+\alpha _{B,0}=-0.8$,
and replace $1+\beta _{B,0}=-2.5$ with $1+\beta _{B,0}=-2$ and
$1+\beta _{B,0}=-5$ respectively to produce the light curve and the
spectral evolutionary curve. In doing so, other parameters remain
unchanged.

\begin{figure}[tbp]
\begin{center}
\includegraphics[width=5in,angle=0]{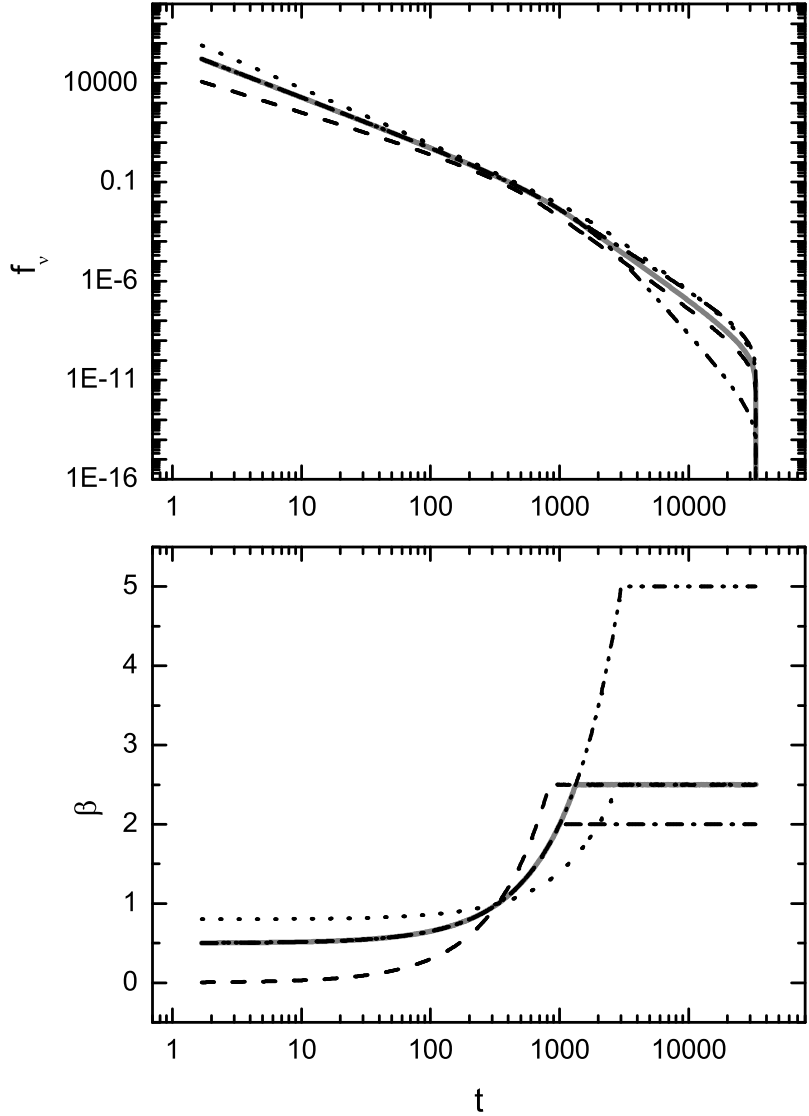}
\end{center}
\caption{The light curve (the upper panel) and spectral evolution
(the lower panel) of the emission considered in Fig. 1 (in the same
units), produced by replacing the indexes adopted there. The dash,
dot, dash dot, and dash dot dot lines correspond to the cases of
replacing $1+\alpha _{B,0}=-0.5$ with $1+\alpha _{B,0}=0$, replacing
$1+\alpha _{B,0}=-0.5$ with $1+\alpha _{B,0}=-0.8$, replacing
$1+\beta _{B,0}=-2.5$ with $1+\beta _{B,0}=-2$, and replacing
$1+\beta _{B,0}=-2.5$ with $1+\beta _{B,0}=-5$, respectively. The
two solid lines in Fig. 1 are presented in gray solid lines.}
\label{Fig. 1_06}
\end{figure}

The results are shown in Fig. 7. As expected, both indexes have
influences on the softening curve: the low energy index $\alpha
_{B,0}$ puts a lower limit to the observed spectral index $\beta$,
making $\beta_{min} = - (1+\alpha _{B,0})$; and the high energy
index $\beta _{B,0}$ confines the upper limit of $\beta$, following
$\beta_{max}=-(1+\beta _{B,0})$. The ratio between the fluxes at
$60s$ and at $3000s$ is also influenced by the indexes, making the
undetectable problem to be eased or worse.

Combining Figs. 5 and 7 we come to the following conclusions: the
observed $\beta_{min}$ and $\beta_{max}$ are determined by the low
and high energy indexes of the observed Band function spectrum (note
that in the case of $\delta$ function emission the observed spectrum
and the intrinsic spectrum share the same form); the observed
$\beta_{min}$ and $\beta_{max}$ for different observation
frequencies would be unchanged as long as the whole softening
process appears within the whole available observation time at the
concerned frequencies.

\section{Application}

\subsection{The light curve and spectral evolution expected in the XRT band}

Our theoretical analysis carried above does not directly correspond
to real observational data. In fact, instead of being defined at a
particular observation frequency, the XRT light curve is measured
within an energy band which is $0.3 - 10$ keV. Therefore, it is
necessary to investigate the light curve as well as the spectral
evolution over this energy range. The available flux of the XRT
light curve is always that has been integrated over this band. We
use $f_{\nu,int}$ to denote this flux which is determined by
\begin{equation}
f_{\nu,int}(t)=\int_{0.3keV}^{10keV}f_{\nu }(t)d\nu.
\end{equation}
The flux is now in units of $erg\cdot cm^{-2} s^{-1}$.

There is a difficulty in evaluating the spectral index over a band.
If the band is large enough one might not be able to consider it
acting still as a power-law. Although we can impose a power-law on
the spectrum over this band, the power-law index is still hard to be
defined and hence hard to be determined. In terms of observation, we
can collect all data points within this band and then figure out the
index by performing a power-law fit. This method is hard to be
adopted in theoretical investigation since one can create countless
data points. We therefor turn to consider a simpler but well defined
approach. First, we assume a power-law over the whole XRT band and
then calculate the index by considering only the fluxes at the lower
(0.3 keV) and upper (10 keV) limits of the band. Second, a power-law
is assumed over a smaller band ($0.6 - 5$ keV) and then the index is
calculated by employing only the fluxes at the new lower (0.6 keV)
and upper (5 keV) limits. Presented in Fig. 8 are the spectral
evolution so evaluated and the light curve of (14), where the Band
function is also adopted and parameters other than the observation
frequency are the same as those adopted in Fig. 1.

\begin{figure}[tbp]
\begin{center}
\includegraphics[width=5in,angle=0]{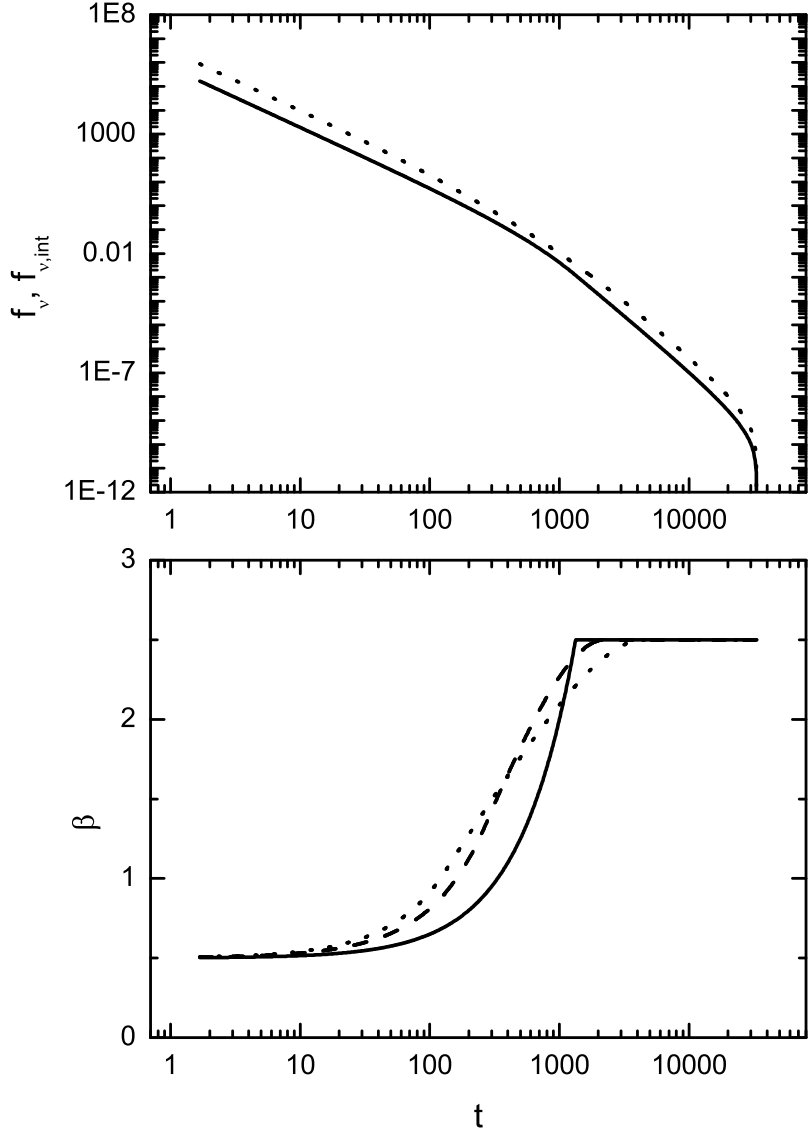}
\end{center}
\caption{The light curve over the XRT band (the dot line in the
upper panel; in unites of $F_0 keV$) and the corresponding spectral
evolution (the lower panel) of the emission considered in Fig. 1.
The dot line in the lower panel represents the spectral index
evaluated by considering a power-law spanning from 0.3 keV to 10
keV; the dash line stand for that calculated by assuming a power-law
spanning from 0.6 keV to 5 keV. For the sake of comparison, both
solid lines in Fig. 1 (in the same units as adopted there) are also
presented.} \label{Fig. 1_01}
\end{figure}

The upper panel of Fig. 8 shows that the XRT light curve is very
similar to the 1 keV light curve. The power-law index $\alpha$ of
the former seems slightly different from that of the latter. This
will lead to a deviation from the $\alpha =2+\beta $ relation. From
the lower panel we find that the lower and upper limits of the
corresponding spectral index ($\beta_{min}$ and $\beta_{max}$) are
the same as that measured at 1 keV, but the spectral evolutionary
curve deviates significantly from that measured at 1 keV. The
deviation is so large that the $\alpha =2+\beta $ relation would be
violently violated. This might help us to understand why this
relation is not commonly detected. The lower panel also shows that
the narrower the power-law range assumed, the closer the spectral
evolutionary curve to that measured at 1 keV. This suggests that,
the narrower band to concern, the more chance of detecting the
$\alpha =2+\beta $ relation.

\subsection{GRB 060614}

Although the condition that the emission is extremely short might be
rare, there might be some bursts their early X-ray emission can
roughly be accounted for by equation (11). Once a burst is selected,
there are two ways of testing. One is to directly fit the light
curve data and the spectral data with equation (11). The other is to
check if their temporal and spectral indexes obey the $\alpha
=2+\beta $ relation. We adopt the second method since the result
does not depend on fitting parameters.

Among the bursts (up to March 28, 2008) analyzed by the UNLV GRB
Group (see http://grb.physics.unlv.edu), GRB 060614 might be one of
such bursts that can be accounted for by the $\delta$ function
emission curvature effect model. There is an obvious softening in
the steep decay phase for this burst. The light curve in this phase
is relatively smooth, suggesting that, besides the main decay
emission, it is unlikely that other components obviously influence
the light curve. In this way, the temporal index can be well
evaluated.

\begin{deluxetable}{cccc}
\tablewidth{0pt} \tablecaption{Temporal indexes of GRB 060614.}
\tablehead{ \colhead{section} & \colhead{$t_1$(s)} &
\colhead{$t_2$(s)} & \colhead{$\alpha$} } \startdata

 1 &   103.3    &   126.8 &   2.73      $\pm$  0.23 \\
 2 &   126.8    &   168.8 &   2.25      $\pm$  0.13 \\
 3 &   169.3    &   309.8 &   3.841      $\pm$  0.064 \\
 4 &   315.0    &   408.4 &   3.78      $\pm$  0.25 \\
 5 &   408.4    &   468.8 &   5.89      $\pm$  0.48 \\
\enddata
\end{deluxetable}

According to the quality of the light curve data of this burst, we
divide them into five sections in this phase, requiring that, for
each section, $log f_{\nu}$ and $log t$ are linearly correlated. We
then estimate the temporal index in each section by fitting the
corresponding data with a power law function. Time intervals of
these sections as well as the fitting results (the estimated
temporal index $\alpha$) are listed in Table 1, where $t_1$ and
$t_2$ are the lower and upper limits of the observation time of the
corresponding sections respectively.

\begin{figure}[tbp]
\begin{center}
\includegraphics[width=5in,angle=0]{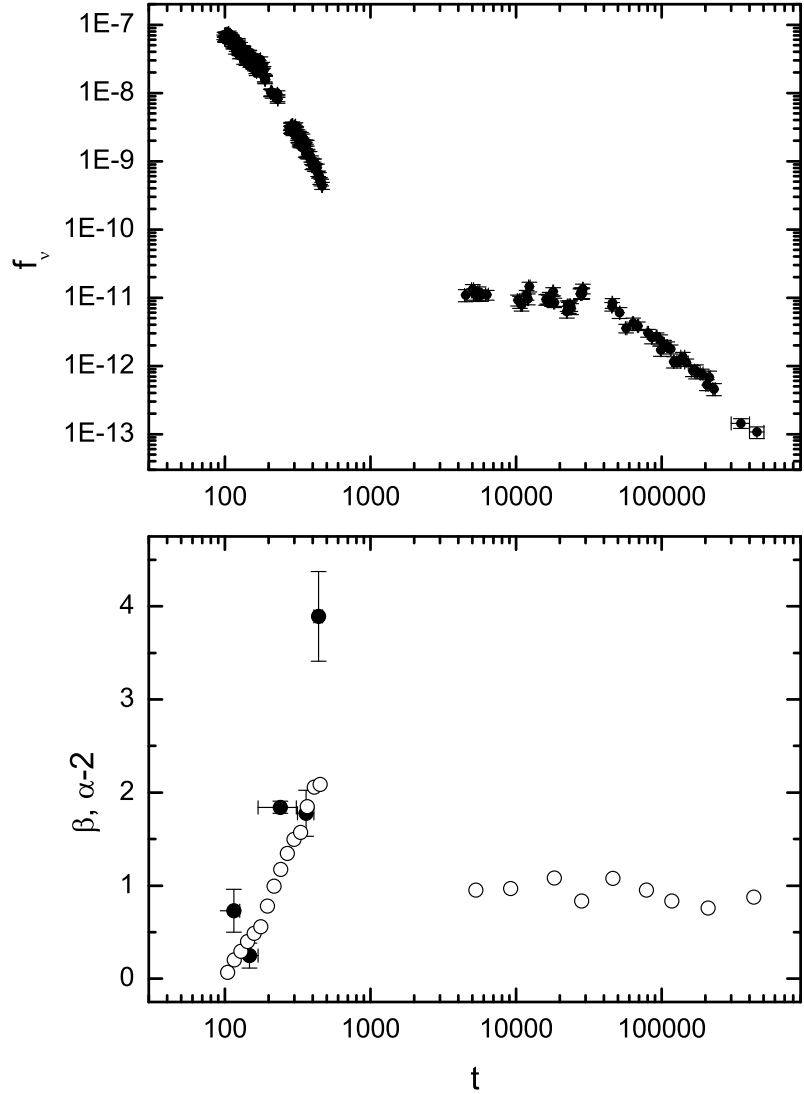}
\end{center}
\caption{Comparison of the temporal index in $\alpha-2$ (the filled
circles in the lower panel) and spectral index $\beta$ (the open
circles in the lower panel) of GRB 060614 in its steep decay phase.
The data of the flux (the upper panel; in units of $erg \cdot
cm^{-2} s^{-1}$) and $\beta$ are taken from the UNLV GRB Group
website.} \label{Fig. 1_07}
\end{figure}

The temporal index obtained from the light curve by the fit and the
spectral index measured in the X-ray band by the UNLV GRB Group are
presented in Fig. 9. We find that the $\alpha -2$ vs. $t$ curve is
roughly in agreement with the $\beta$ vs. $t$ curve in the concerned
steep decay phase, suggesting that the curvature effect might
probably be the main cause of the steep decay curve.

As analyzed in last subsection, the $\alpha =2+\beta $ relation will
not be strictly obeyed if one considers the light curve and the
spectral index over a band (here, the XRT band) instead of at a
fixed frequency. A deviation between the $\alpha -2$ vs. $t$ curve
and the $\beta$ vs. $t$ curve is hence expectable. However, the
temporal index measured in the last time section in this phase is so
large that it is likely to have other causes. Although the effect of
the light curve over a band and the effect of the duration of the
intrinsic emission have not been considered, we still suspect that
the large temporal index might be a consequence of larger absorption
for higher latitude photons.

\section{Discussion and conclusions}

We investigate the influence of the curvature effect on both the
light curve and the spectrum of late emission of GRBs, attempting to
explain the observed softening phenomenon in early X-ray afterglows
of Swift. As the first step of investigation, we explore only the
case of extremely short intrinsic emission, for which we assume and
apply a $\delta$ function emission. Although how an emission is
extremely short is currently unclear (this deserves a detailed
investigation in the near future) and the condition that the
emission is extremely short might be rare, the investigation is
necessary since by considering such emission the possible impacts
from the emission duration can be ignored and then the pure
curvature effect can plainly be illustrated.

Formulas presented in Qin (2008) are employed to study this issue.
Unlike what investigated in Qin (2008) and other relevant
theoretical analyses (e.g., Fenimore et al. 1996; Sari et al. 1998;
Kumar \& Panaitescu 2000), we do not limit our study on an intrinsic
power law spectrum. Instead, we consider more general spectral form
of emission. Assuming a $\delta$ function emission we get formulas
that get rid of the impacts from the intrinsic emission duration,
which are applicable to any forms of spectrum. According to these
formulas, one would observe the same form of spectrum at different
times, with the peak energy of the spectrum shifting from higher
energy bands to lower bands. This was detected recently in the early
X-ray afterglows of some GRBs (see Butler \& Kocevski 2007). The
peak energy is expected to decline following $E_{peak}\propto
t^{-1}$. In the case when the emission is early enough so that the
emitting area on the fireball surface is not far from the line of
sight (say, $\theta \ll \pi/2$ or $t \ll R_{c}/v$), the temporal
power law index and the spectral power law index will be well
related by $\alpha =2+\beta $. As a consequence, the form
$t^{2}f_{\nu }(t)$ will possess exactly the intrinsic spectral form
(say, when replacing $t$ with $\nu$ one will get from $t^{2}f_{\nu
}(t)$ as a function of $\nu$ that take exactly the same form of the
intrinsic spectrum). The $\alpha =2+\beta $ relation will be broken
down and $\alpha
> 2+\beta $ or $\alpha \gg 2+\beta $ will hold (see Fig. 4) at much later time when
the angle between the moving direction of the emission area and the
line of sight is large.

As revealed in Butler \& Kocevski (2007) and suggested by Starling
et al. (2008), we focus our attention to the emission with a Band
function spectrum. This spectrum has a power law behavior in both
lower and higher energy bands, where the two power laws are smoothly
connected (Band et al. 1993). Using this spectral form, we plot the
light curve and the spectral evolution with our formulas, expected
at 1 keV and other frequencies. The analysis shows that there do
exist a temporal steep decay phase and a spectral softening which
occur simultaneously. As Fig.2 reveals, both the steep decay light
curve and the spectral softening are caused by the shifting of the
Band spectrum. As mentioned above, Starling et al. (2008) suggested
that the steep decline in the flux of GRB 070616 may be caused by a
combination of the strong spectral evolution and the curvature
effect. Based on the above argument, we insist that both the
spectral evolution and the steep decline observed in GRB 070616 and
other Swift bursts are likely to be caused merely by the curvature
effect. In addition, we find that, just as what is illustrated in
Qin (2008), the $cos\theta$ term, $R_{c}/v-t$, plays a role in
producing the light curve, which ``attaches'' a cutoff tail to the
latter (see Fig. 1). The spectral softening terminates when the
emission is dominated by that from the high energy portion of the
Band function spectrum. Thus, there exists a maximum of the spectral
index, $\beta_{max}$. After the $\beta_{max}$ appears, it lasts to
the end of the emission, including the phase of the cutoff tail. Our
analysis shows that the softening due to the curvature effect will
appear at different frequencies; it occurs earlier for higher
frequencies and later for lower frequencies; a characteristic of the
softening, the terminating softening time $t_{s,max}$ (when the
$\beta_{max}$ appears), depends strictly on the observation
frequency, which follows $t_{s,max}\propto \nu^{-1}$. Although this
is concluded based on the assumption of extremely short emission, we
tend to believe that its main characters hold in most cases since
the light curve from any finite emission is contributed by countless
$\delta$ function emission. The combination of these countless
$\delta$ function emission would change the values of some
quantities such as $t_{s,max}$, but would not change the trend.
Starling et al. (2008) showed, the spectral evolution of GRB 070616
starts earlier at $ \gamma$-ray energies (while the X-ray flux is
still at an approximately constant level) and begins much later at
X-ray energies around the onset of the steep X-ray decay. In terms
of the curvature effect, this is due to the following fact: the peak
energy of the Band function spectrum passes through the $\gamma$-ray
band at a relatively early time, while it passes through the X-ray
band at a later time.

Parameters that might have impacts on the light curve and the
spectral evolution are also studied. Whilst a smaller Lorentz factor
shifts the spectral softening to larger time scales, smaller values
of the fireball radius and the rest frame peak energy make the
occurrence of the softening earlier. The analysis shows that the
terminating softening time appears much later than the start time of
the softening. The former is always larger than the latter by about
one order of magnitude. It is therefore predicted that the duration
of the softening and the terminating softening time would be
linearly correlated. It also shows that the low energy index $\alpha
_{B,0}$ puts a lower limit to the observed spectral index $\beta$
and the high energy index $\beta _{B,0}$ confines the upper limit of
$\beta$. The following conclusions are reached: the observed
$\beta_{min}$ and $\beta_{max}$ are determined by the low and high
energy indexes of the observed Band function spectrum; $\beta_{min}$
and $\beta_{max}$ for different observation frequencies would remain
unchanged as long as the whole softening process appears within the
whole available observation time at the concerned frequencies.

As application, we study the light curve and the spectral evolution
over the XRT band. That is, the light curve is that has been
integrated over the $0.3-10$ keV band and the spectral index is
estimated by assuming a power-law over this band. The analysis shows
that the light curve slightly deviate from that measured at 1 keV,
whilst the spectral evolutionary curve significantly betrays that
measured at 1 keV. This suggests that the $\alpha =2+\beta $
relation will be violently violated if one measures the light curve
and estimates the spectral index over a wide band.

Another application is to check the $\alpha =2+\beta $ relation by
employing the light curve and spectrum data of GRB 060614. The
temporal index $\alpha$ in the steep decay phase of this burst is
evaluated. We compare $\alpha$ and the spectral index $\beta$ in the
same softening phase (it is also the steep decay phase of the light
curve). It shows that the $\alpha-2$ vs. $t$ curve and the $\beta$
vs. $t$ curve are roughly in agreement, suggesting that the $\alpha
=2+\beta $ relation is roughly satisfied in this phase and the
softening of this burst might possibly be due to the curvature
effect.

What we have investigated are based on the assumption of extremely
short emission which ignores the possible impacts from the emission
duration. While the contribution of the emission duration might have
less affect on the spectrum, it might obviously affect the temporal
profile and hence the temporal index. Therefore the $\alpha =2+\beta
$ relation might not hold in this case. Combining this with the
problem arising from the estimation of the spectral index in the
softening phase over an energy band rather than at a fixed
frequency, it might be able to explain why the $\alpha=2+\beta$
relation is not common in Swift bursts (see Liang et al. 2006 and
also Zhang 2007 for a detailed discussion). Yonetoku et al. (2008)
showed that the two characteristic break energies they considered
have a time dependence of $ \propto t^{-3}-t^{-4}$. This is not in
agreement with the prediction of the $E_{peak}\propto t^{-1}$ law.
Perhaps an intrinsic softening might be responsible for this
difference when the emission duration is taken into account. Another
possibility is that the $E_{peak}$ shifting of this kind of burst is
due to structure jets, where in high latitude, $E_{peak}$ would be
much smaller than it is in uniform jets. In addition to test the
$\alpha =2+\beta $ relation and the $E_{peak}\propto t^{-1}$ law, we
suggest to detect the softening at different frequencies as well.
The observed relations might not strictly follow what the $\delta$
function emission model predicts, but as argued above, the trend of
the relevant effects would be maintained (which deserves a detailed
investigation in the near future).

\begin{figure}[tbp]
\begin{center}
\includegraphics[width=5in,angle=0]{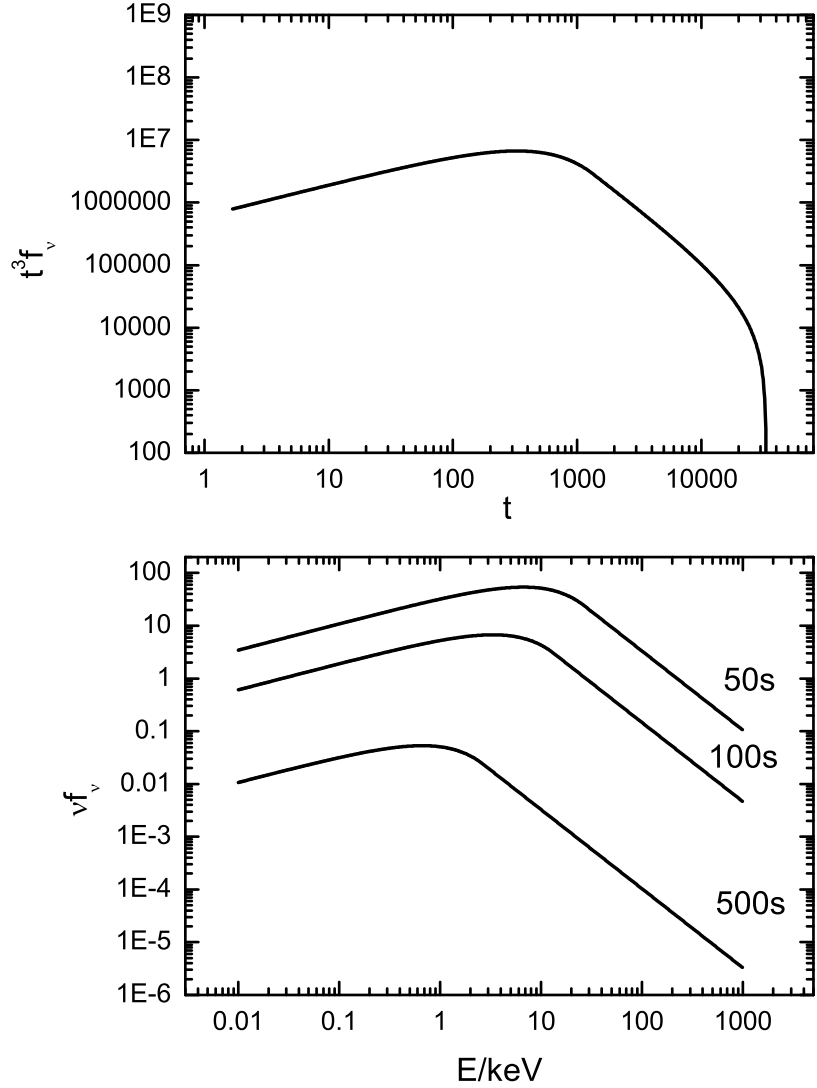}
\end{center}
\caption{Comparison of the functions of the $t^{3}f_{\nu }(t)$ vs.
$t$ curve (the upper panel) and the $\nu f_{\nu}$ vs. $\nu$ curve
(the spectrum; the lower panel) of the emission considered in Fig.
1. The three solid lines in the lower panel are the same lines of
50, 100, and 500 s respectively in Fig. 1.} \label{Fig. 1_08}
\end{figure}

In addition to these more conventional tests, we propose to try a
totally new test which is to check if the $t^{3}f_{\nu }(t)$ vs. $t$
curve is in agreement with the $\nu f_{\nu}$ vs. $\nu$ curve when
replacing $t$ with $\nu$ in the $t^{3}f_{\nu }(t)$ and $t$ forms and
multiplying a constant to match the corresponding dimensions (see
Section 3). Illustrated in Fig. 10 is an example of the comparison
(where we compare only the relevant functions and thus do not
replace variables or multiply constants to change the dimensions).
An advantage of doing so is that we can guess spectrum form merely
from the light curve data. For example, when we find that the
$t^{3}f_{\nu }(t)$ form being a perfect power-law function of time,
then the spectrum is guessed to be a pure power-law (if the spectral
data are found not to obey a power-law, then we will have reasons to
doubt if this burst is not affected by the curvature effect). Or,
when we find that the $t^{3}f_{\nu }(t)$ form is a Band function of
time, then we will have reasons to guess that the spectrum takes a
Band function form. Another usage of plotting the $t^{3}f_{\nu }(t)$
vs. $t$ curve is to find out the time when the peak energy passes
through the observation band (se also Kumar \& Panaitescu 2000),
which can be directly checked by observation and hence becomes a
test to the curvature effect as well. According to equation (11),
the moment when the peak of $t^{3}f_{\nu }(t)$ appears is the time
when the peak energy passes through the observation frequency $\nu$,
or, it is the time when $E_{peak}=h\nu$. In plotting the
$t^{3}f_{\nu }(t)$ vs. $t$ curve (see the upper panel of Fig. 10),
one can also check the influence of factors other than that of the
pure curvature effect by observing and measuring the deviation of a
real light curve from that of equation (11). An extra usage of this
plot might be a direct comparison of light curves of different
bursts, probably enabling us to divide them according to their
temporal properties (we strongly suggest a detailed investigation on
this issue in the near future).

\begin{figure}[tbp]
\begin{center}
\includegraphics[width=5in,angle=0]{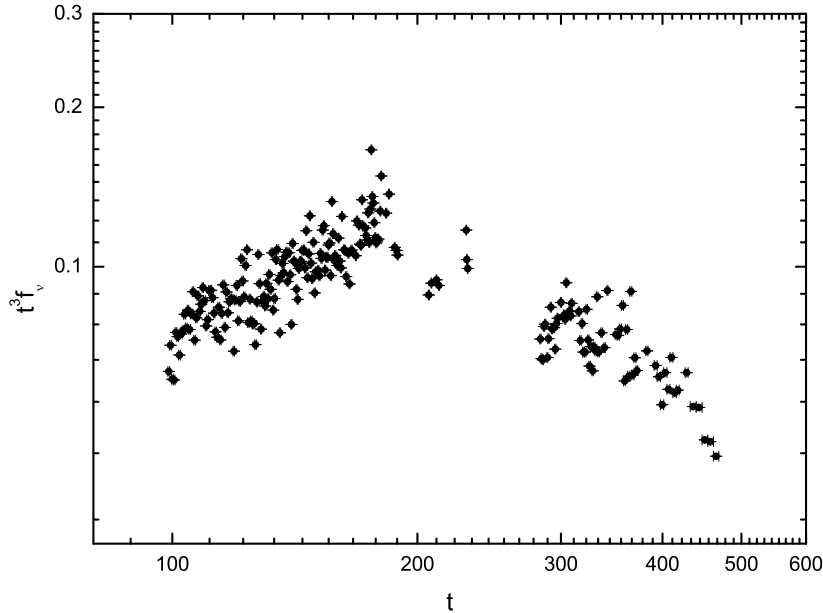}
\end{center}
\caption{The $t^{3}f_{\nu }(t)$ (in units of $erg \cdot cm^{-2}
s^{2}$) vs. $t$ curve of GRB 060614 in its steep decay phase. The
data are taken from the UNLV GRB Group website.} \label{Fig. 1_09}
\end{figure}

Displayed in Fig. 11 is the $t^{3}f_{\nu }(t)$ vs. $t$ curve of GRB
060614. It shows a function of time bearing the Band function form
(when replacing $\nu$ with $t$) attached with a cutoff tail. If we
believe that the softening of this burst is due to the curvature
effect in the case of extremely short emission, according to Fig. 11
it would be expectable that the peak energy of this burst passes
through the corresponding observation energy range (the XRT band) at
$\sim $175 s (or, at $\sim $175 s, the peak energy of the observed
spectrum is just within the energy range of the light curve).
Indeed, as presented in Table 4 of Mangano et al. (2007), within the
time interval $128-190$ s, the peak energy (when fitting the
spectrum with a Band model) is about $2-4$ keV which is well within
the $0.3-10$ keV band. A similar result for this burst is also
visible in Fig. 1 of Butler \& Kocevski (2007).

Based on their fits to the composite light curves, Sakamoto et al.
(2007) confirmed the existence of an exponential decay component
which smoothly connects the BAT prompt data to the XRT steep decay
for several GRBs. Yonetoku et al. (2008) also showed that the
spectrum of GRB 060904A contains a cutoff tail in its higher energy
range, which can be represented by an exponential function.
According to the above analysis, the spectral form obviously affects
the light curve if the curvature effect is at work and the intrinsic
emission is short enough. An intrinsic spectrum with an exponential
tail might probably lead to an exponential decay light curve. We
notice that the projected factor (the $\cos \theta $ term) also
produces a very steep decay phase when the angle between the moving
direction of the dominant emission area and the line of sight is
large enough. However, it is unlikely that many of them (if not only
few of them) are due to the $cos\theta$ term, since this term always
appears at a very late time (see Fig. 6 and also Qin 2008). The
following factors might also be the cause of this tail: one is the
large absorption for higher latitude emission and the other is the
limited open angle of jets. Both will lead to a steeper tail. While
the former might probably give rise to a smooth decay curve, the
latter might probably lead to a sharp feature.

In addition to the shifting of the peak energy, Starling et al.
(2008) also observed softening of the low energy power law slope.
They measured a softening of the low energy spectral slope from
$\alpha \sim 0.1 - 1.3$. This implies that the intrinsic spectrum
might evolved itself. This cannot be taken into account in the
current investigation since a $\delta$ function emission has no
evolution. We suggest to explore in a later investigation the impact
from the emission duration and the effect from the intrinsic
spectral evolution.

\acknowledgments

Special thanks are given to the anonymous referee for his or her
comments and suggestions which have improved the paper greatly. This
work was supported by the National Natural Science Foundation of
China (No. 10573005 and No. 10747001) and by the Guangzhou Education
Bureau and Guangzhou Science and Technology Bureau.

\appendix

\section{Projected factor as a function of time}

Emission from a distant area is proportional to the projected factor
of the area relative to the observer, which is known as $cos\theta$,
where $\theta$ is the angle between the normal of the area and the
line of sight. For a face-on area (its normal is parallel to the
line sight), $\theta=0$ and then $cos\theta=1$; and for an edge-on
area (its normal is perpendicular to the line sight), $\theta=\pi/2$
and then $cos\theta=0$. The projected factor varies from
$cos\theta=1$ to $cos\theta=0$ for the half fireball surface facing
the observer. Due to time delation, emission from the area of
$\theta=0$ reaches the observer earlier and that of $\theta=\pi/2$
reaches the observer later, and the former emission is not reduced
by the projected factor whilst the latter will be reduced to zero.
It has been pointed out that the term $R_{c}/v-t$ in equation (11)
comes from nothing but the projected factor $cos\theta$ (Qin 2008).
Here we analyze this issue in detail.

Taking into account the increase of the radius of an expanding
fireball (equation A.3 in Qin 2002) and the time delay associated
with different latitudes in the fireball surface (equation A.6 in
Qin 2002), one obtains the following relation between the
observation time $t_{ob}$ and the angle relative to the line of
sight (equation A.7 in Qin 2002)
\begin{equation}
t_{emit}=\frac{t_{ob}-D/c+[R_{c}/c-(v/c)t_{c}]cos\theta}{1-
(v/c)cos\theta },
\end{equation}
which could be written as
\begin{equation}
cos\theta=\frac{t_{emit}+D/c-t_{ob}}{R_{c}/c+(v/c)(t_{emit}-t_{c})}.
\end{equation}
Replacing $t_{ob}$ with $t$ (see equation 1) and considering the
time contraction $t_{emit}-t_c=\Gamma (t_0-t_{0,c})$ (equation A.1
in Qin 2002), we get
\begin{equation}
cos\theta=\frac{(t_0-t_{0,c})\Gamma +R_c/v-t}{[(t_0-t_{0,c})v\Gamma
+R_{c}]/c}.
\end{equation}
As revealed in equation (A.4) in Qin (2002), $(t_0-t_{0,c})v\Gamma
+R_{c}$ is the fireball radius measured at $t_0$. Equation (A3)
suggests that, for any particular intrinsic emission time $t_0$ (at
which the fireball radius is fixed), $cos\theta$ varies with
observation time $t$. This is due to the time delay, as mentioned
above (see also equation A.6 in Qin 2002). According to equation
(A3), the term $(t_0-t_{0,c})\Gamma +R_c/v-t$ in equation (2) is the
projected factor $cos\theta$ multiplying the fireball radius divided
by the speed of light. Note that the term $(t_0-t_{0,c})\Gamma
+R_c/v-t$ in equation (2) directly comes from the term $cos\theta$
in equation (A.15) in Qin (2002) as a consequence of the projected
effect (see equation A.11 in Qin 2002).

For the $\delta$ function emission considered in this paper, the
intrinsic emission is assigned to occur at $t_0=t_{0,c}$ (see
equation 7), which leads to
\begin{equation}
cos\theta=\frac{R_c/v-t}{R_{c}/c}.
\end{equation}
This explains why we interpret the term $R_c/v-t$ in equation (11)
as that reflecting the projected factor of the emission area in the
fireball surface.

Note that observation time $t$ is confined by equation (8) which
arises from the limit of the fireball surface in the case of
$\delta$ function emission (see Qin 2002; Qin et al. 2004; Qin
2008). According to equation (A4), $cos\theta=1$ when
$t=(1-v/c)R_c/v$ which is the observation time when photons from the
face-on area ($\theta=0$) of the fireball surface arrive the
observer. When taking $t=R_c/v$ we get $cos\theta=0$. Note that
$t=R_c/v$ is the observation time when photons from the edge-on area
($\theta=\pi/2$) reach the observer.

In the case of $R_c=10^{15}cm$ and $\Gamma=100$, $R_c/v \sim 3\times
10^4 s$ which is much larger than the steep decay segment of most
GRBs detected by Swift. Even for $R_c=10^{14}cm$ we get $R_c/v \sim
3\times 10^3 s$ which is still larger than the steep decay segment
of many Swift GRBs. For the observation time satisfying $t\ll
R_c/v$, the projected factor approaches a unit and then can be
ignored. Only when the observation time is comparative to $R_c/v$,
the projected factor will play an important role (see the broken
down feature in Fig. 1). The broken down feature also exists in the
case of the pure power-law spectrum (Qin 2008). When this feature is
observed, the fireball radius will be well determined, no matter the
observed spectrum is a pure power-law or a Band function form.

\section{Projected factor in flux}

Here we show how the projected factor plays a role in the flux per
unit frequency from a relativistic source.

As shown in Qin (2002), the amount of energy emitted from
differential area $ds_{emit}$ radiating towards the observer is
\begin{equation}
dE_{emit}=\frac{I_{\nu }cos\theta ds_{emit}d\nu dt \Delta
s_{ob}}{D^2},
\end{equation}
where $I_{\nu }$ is the intensity of radiation measured by local
observers near the emitting source, $cos\theta$ is the projected
factor of $ds_{emit}$ towards the distant observer, $d\nu$ and $dt$
are the observed frequency and time intervals respectively, $\Delta
s_{ob}$ is the area of the detector, and $D$ is the distance between
the observer and the emitter. The flux density measured from this
amount of energy is
\begin{equation}
d f_{\nu}=\frac{I_{\nu }cos\theta ds_{emit}}{D^2}.
\end{equation}

Observer frame intensity $I_{\nu }$ is related with the intrinsic
intensity $I_{0,\nu }$ by
\begin{equation}
I_{\nu }=(\frac{\nu}{\nu_0})^3I_{0,\nu },
\end{equation}
where $\nu_0$ is the intrinsic frequency. Applying this relation and
the Doppler effect we get
\begin{equation}
d f_{\nu}=\frac{I_{0,\nu }cos\theta ds_{emit}}{D^2\Gamma^3
[1-(v/c)cos\theta]^3}.
\end{equation}
If the area is on the tip of the fireball surface, $\theta=0$. That
gives rise to
\begin{equation}
d f_{\nu,0}=\frac{I_{0,\nu } ds_{emit}}{D^2\Gamma^3 [1-(v/c)]^3}.
\end{equation}
The relation between $d f_{\nu}$ and $d f_{\nu,0}$ is
\begin{equation}
\frac{d f_{\nu}}{d
f_{\nu,0}}=[\frac{1-(v/c)}{1-(v/c)cos\theta}]^3cos\theta.
\end{equation}

For a large Lorentz factor, we get
\begin{equation}
\frac{d f_{\nu}}{d
f_{\nu,0}}=\frac{cos\theta}{8\Gamma^6[1-(1-1/2\Gamma^2)cos\theta]^3}.
\end{equation}
Here are several particular results: a) when $\theta \ll 1/\Gamma$,
$d f_{\nu}/d f_{\nu,0}=1$; b) when $1/\Gamma \ll \theta \ll 1$, $d
f_{\nu}/d f_{\nu,0}=1/(\theta\Gamma)^{6}$, which was previously
pointed out by Kumar \& Panaitescu (2000); c) when $\theta=\pi/3$,
$d f_{\nu}/d f_{\nu,0}=1/2\Gamma^{6}$; d) when $\theta=\pi/2$, $d
f_{\nu}/d f_{\nu,0}=0$.

\end{document}